\documentclass[aps,twocolumn,prb]{revtex4-1}
\usepackage{amsfonts}
\usepackage[final,hiresbb]{graphicx}
\usepackage{amsmath}
\usepackage{amssymb}
\usepackage{amstext}
\usepackage{braket}
\usepackage[colorlinks,allcolors=blue]{hyperref}

\usepackage[mathscr]{euscript}

\usepackage{color}

\definecolor{Green}{RGB}{0,204,102}
\definecolor{Purple}{RGB}{102,0,255}
\definecolor{Blue}{RGB}{51,153,255}
\definecolor{Red}{RGB}{151,010,010}

\begin{document}

\title{Designing Small Silicon Quantum Dots with Low Reorganization Energy}  
\author{Xiaoning Zang and Mark T. Lusk}
\email{mlusk@mines.edu}
\affiliation{Department of Physics, Colorado School of Mines, Golden, CO 80401, USA}

\begin{abstract}
A first principles, excited state analysis is carried out to identify ways of producing silicon quantum dots with low excitonic reorganization energy.  These focus on the general strategy of either reducing or constraining exciton-phonon coupling, and four approaches are explored. The results can be implemented in quantum dot solids to mitigate polaronic effects and increase the lifetime of coherent excitonic superpositions. It is demonstrated that such designs can also be used to alter the shape of the spectral density for reorganization so as to reduce the rates of both decoherence and dissipation. The results suggest that it may be possible to design quantum dot solids that support partially coherent exciton transport.  
\end{abstract}
\maketitle

%
\section{Introduction}
Recent progress in understanding the electronic wave functions of colloidal semiconductor quantum dots (QDs) has led to an ability to synthesize isolated, quantum confined building blocks with a variety of tailored optical properties. Quantum dots composed of silicon (SiQDs) seem particularly promising going forward for many biomedical, display, computing and solar energy applications. This is because they are environmentally benign, resource plentiful and benefit from decades of industrial know-how. SiQDs have been shown to carry out multiple-exciton generation (MEG), \cite{Beard13, Timmerman16, Timmerman17} efficiently transport excitons,\cite{Lin19} have higher coupling for same surface-to-surface separation,\cite{Lin19} are free of defects\cite{Niesar20} and resist oxidation better\cite{Li21}; SiQDs encapsulated within an inorganic amorphous matrix\cite{Conibeer22, Conibeer23, Hao24} and within an organic polymer blend\cite{Herrmann25, liu26, AADT27, Gowrishankar28, Svrcek29, Liu30, Niesar31, Li32} for photovoltaic application has been studied. 

For many applications, such as light emission,\cite{Cheng2011} optical computing,\cite{Daldosso2009}, and biomedical imaging, sensing and treatments,\cite{LiuBio2013, Peng2014, Tomczak2013} carrier transport is not a primary concern. Within the solar energy arena, though, quantum dots solids need to exhibit efficient carrier dynamics.\cite{Alivisatos10, Nozik11, Jurbergs12, Beard13, Madrid14, Nozik15, Timmerman16, Timmerman17, Lin18} Transport of energy and charge is now a central issue in realizing the mesoscopic potential of quantum dot solids. This is a critical bottleneck because charge and exciton transport tend to proceed via low mobility, incoherent hopping associated with polaronic trapping and weak electronic coupling.

For quantum dot solids in which excitons readily dissociate, a number of promising strategies seek to improve charge transport by focusing on properties such as translational symmetry, electronic overlap, matrix encapsulation, and crystalline orientation. Dynamics within materials composed of small SiQDs, though, tend to be dominated by the motion of excitons though. While quantum confinement in SiQDs offers benefits, such as a pseudo-direct gap and high emission cross section, it also results in a low dielectric constant. This implies low screening and a high excitonic binding energy so that assemblies of SiQDs support excitons with a Frenkel character. Their dissociation is sufficiently problematic to motivate a consideration of ways in which the excitons themselves can be used to efficiently conduct energy. This is reminiscent of exciton transport that is efficiently carried out in photosynthetic complexes, offering either biomimetic or bio-inspired solution strategies.

\begin{figure}[hptb]
\begin{center}
\includegraphics[width=0.48\textwidth]{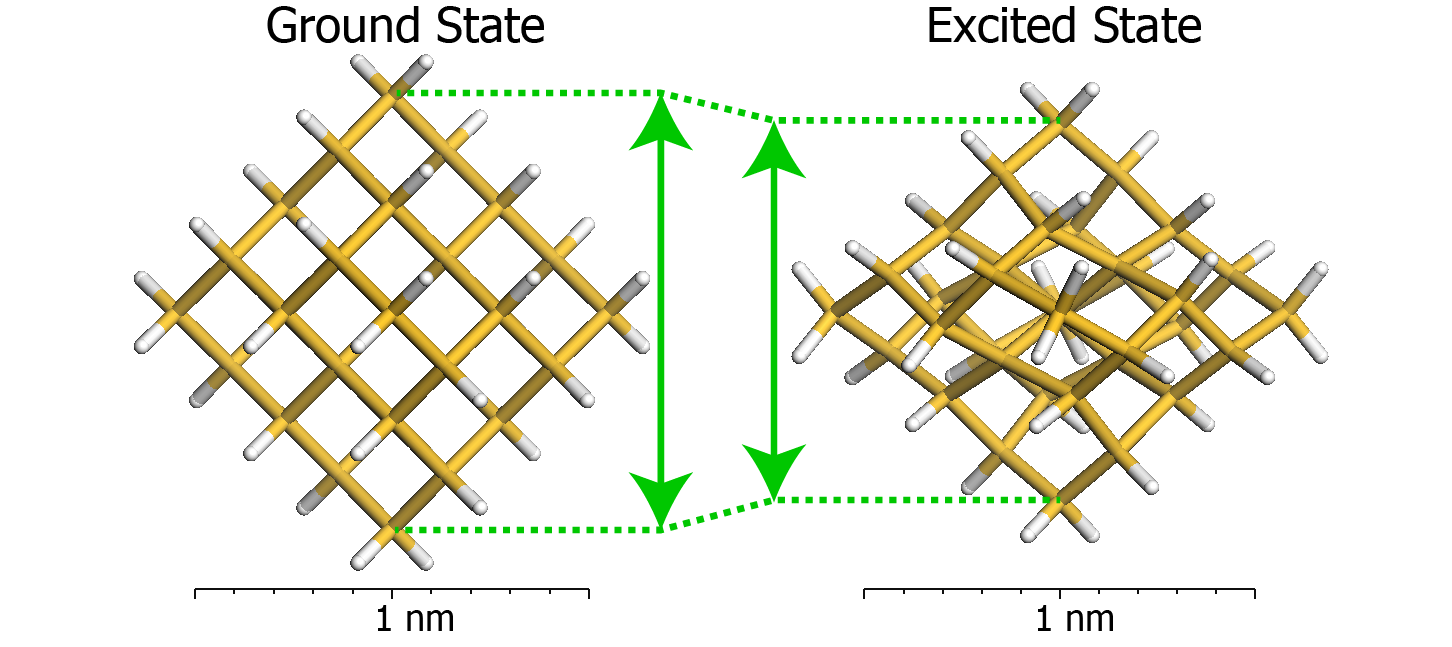}
\end{center}
\caption{Reorganization following QD excitation for $\mathrm{Si}_{35}\mathrm{H}_{36}$. Ground state structure (left) relaxes into a new structure (right) when promoted into one of its degenerate first excited singlet states.  As a result, the dot contracts along the vertical axis. The displacement is magnified 5 times in order to make this distortion clear.}\label{distort}
\end{figure}

In an ideal assembly of such dots, photon absorption results in a coherent superposition of Frenkel excitons that will subsequently exhibit ballistic travel through the quantum dot solid. In reality, though, such coherence and the associated wavelike transport is rapidly lost. This is because the photoexcitation of a quantum dot shifts its equilibrium geometry and the dot shape will reorganize into a new configuration, as shown in Figure \ref{distort}. The energy change associated with such a relaxation is a measure of exciton-phonon entanglement known as the reorganization energy, $\lambda$. Such exciton-photon entanglement destroys the quantum phase coherence among the excitonic superposition, and transport reduces to a diffusive random walk of excitonic states with a statistical distribution of occupation probabilities. 

%

In the absence of any structural disorder, this reduction to diffusive motion results in a slower rate of excitonic transport. The coherent wave packet, though, can quickly become trapped in naturally occurring disordered regions\cite{Anderson}, so that the reduction to diffusive transport may improve the overall transport rate in realistic QD solids. 

Intriguingly, it may be possible to combine the best of both transport regimes so that a superposition of Frenkel excitons moves in a {\it partially coherent} manner. Inspiration for such a goal comes from photosynthetic complexes\cite{Fenna2, Li3} wherein proteins and pigments are exquisitely structured so the rate of phonon entanglement is slow and the hopping transport is high--prerequisites for exciton transport with a degree of wave-like character\cite{Engel1, Adolphs4, Vulto5,Cho6, Brixner7}. At issue is whether or not such a balance can be achieved in QD solids. 

\begin{figure}[hptb]
\begin{center}
\includegraphics[width=0.5\textwidth]{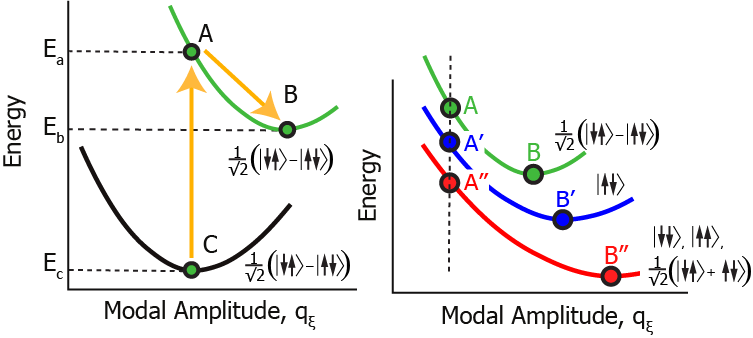}
\end{center}
\caption{ (Left) Vertical excitation to an excited state manifold, due to photon absorption, is followed by reorganization of the nuclei. The overall displacement can be spectrally resolved into that of individual phonon mode amplitudes, $q_\xi$. The excited state manifold is assumed to be a displaced version of its ground state counterpart.(Right) Excited state surfaces depend on spin state as depicted for the pair of orbitals comprising an exciton.}
\label{PES}
\end{figure}

First consider the Hamiltonian of an isolated quantum dot, restricted to be in either its ground state, referenced to zero energy, or first excited singlet state with energy, $\varepsilon_0$. If $\hat Q_\xi$ is the dimensional ($\rm{length}\times \sqrt{mass}$) amplitude operator of vibrational mode $\xi$ with associated momentum operator $\hat P_\xi$, their non-dimensional counterparts are given by $\hat q_\xi = \hat Q_\xi \sqrt{\frac{\omega_\xi}{\hbar}}$ and  $\hat p_\xi = \frac{1}{\sqrt{\hbar\omega_\xi}} \hat P_\xi$. The Hamiltonian is then
\begin{equation}
\hat H^{\rm (dot)} = \sum_{\xi} \frac{\hbar\omega_\xi }{2}\bigl(\hat p^2_\xi +  \hat{c}\hat{c}^{\dagger}\hat q_\xi ^2 \bigr)+\hat{c}^{\dagger}\hat{c} \bigl( \varepsilon_0 +\sum_{\xi} \frac{\hbar\omega_\xi }{2}( \hat q_\xi - \hat q_{0\xi})^2 \bigr) 
\end{equation}
where $\hat c$ is the fermionic exciton annihilation operator and the equilibrium amplitude operator of each vibrational mode is $\hat q_{0\xi}$. This can be re-grouped and simplified, using the bosonic phonon annihilation operator, $\hat b_{\xi} = (\hat q_{\xi }+i \hat p_{\xi })/\sqrt{2}$, into components associated with excitonic site energy, $\hat H_{X}$, phonon energy, $\hat H_{\mathrm{ph}}$, reorganization energy, $\hat H_{\mathrm{reorg}}$, and exciton-photon coupling energy, $\hat H_{\mathrm{X\mbox{-}ph}}$:
\begin{equation}
\hat H^{\rm (dot)}=\hat H_{X} +  \hat H_{\mathrm{ph}} + \hat H_{\mathrm{reorg}} +\hat H_{\mathrm{X\mbox{-}ph}}
\end{equation}
where
\begin{eqnarray}
&&\hat H_{\mathrm{X}}=\varepsilon_0\hat{c}^{\dagger}\hat{c} \nonumber \\
&&\hat H_{\mathrm{ph}}=\sum_{\xi}\hbar \omega_\xi\bigl(\hat{b}_{\xi}^{\dagger}\hat{b}_{\xi}+\frac{1}{2}\bigr) \\
&&\hat H_{\mathrm{reorg}}=\frac{1}{2}\sum_{\xi}\hbar\omega_\xi \hat q^2_{0\xi}\hat{c}^{\dagger}\hat{c} \\
&&\hat H_{\mathrm{X\mbox{-}ph}}=-\frac{1}{\sqrt{2}}\hat{c}^{\dagger}\hat{c}\sum_{\xi}\hbar \omega_\xi \hat q_{0\xi}(\hat{b}_{\xi}^{\dagger}+\hat{b}_{\xi}) \nonumber
\end{eqnarray}
The reorganization energy (Figure \ref{PES}) in the position basis is therefore given by $\lambda :=  \frac{1}{2}\sum_{\xi}\hbar\omega_\xi  q^2_{0\xi} \equiv \sum_{\xi}\lambda_{\xi}$. Huang-Rhys factors are obtained by dividing its spectrally resolved components, $\lambda_{\xi}$, by $\hbar\omega_{\xi}$.  
 %
%
This additive decomposition lends itself to a continuum description of the exciton-phonon coupling as a smooth function of vibrational frequency,
\begin{equation}
\mathscr{J}(\omega)=2\pi\sum_{\xi}\hbar \omega_\xi  \lambda_{\xi}\delta(\omega - \omega_\xi ),
\label{specdens0}
\end{equation}
known as the \emph{spectral density} function. A broadened version of this function is approximated as a superposition of shifted Drude-Lorentz distributions:\cite{Kreisbeck2012, Kreisbeck2013}
\begin{equation}
\mathscr{J}(\omega)=\sum_{i=1}^N\frac{2 \lambda_i \gamma_i \omega}{\gamma_i^2+(\omega-\Omega_i)^2},
\label{specdens}
\end{equation}
where $\gamma_i$ and $\Omega_i$ characterize the width and shift of the $i^{\mathrm{th}}$ Drude-Lorentz peak respectively. The rate at which structural reorganization occurs following excitation can, in large measure, be anticipated by identifying two features of this distribution. This will be considered after constructing a Hamiltonian that allows for exciton transport through an assembly of quantum dots.

As a computational expedient, each dot within an assembly can be treated as a lattice site, $n$, within discrete meso-scale Hamiltonian with a subscript:\cite{Yang, Chris}
\begin{equation}
\hat H=\sum_n \hat H^{\rm (dot)}_n + \sum_{m\neq n}J_{m n}\hat c^{\dagger}_m\hat c_n 
\end{equation}
Here $J_{mn}$ is the coupling between the $m\mathrm{th}$ and $n\mathrm{th}$ sites. The character of exciton transport  can be then delineated by the size of these coupling parameters relative to the associated reorganization energies. Suppose all of the dots and and their couplings to be identical so that the assembly is characterized by single $\lambda$ and $J$ values. Then $\lambda \ll$ J leads to the the coherent transport dynamics of the Redfield equation\cite{KuhnAndMay, Ishizaki2} and $\lambda \gg$ J results in the incoherent, hopping transport of the F$\mathrm{\ddot{o}}$rster equation \cite{Foster1, Foster2}. The intermediate case, wherein $\lambda \sim$ J, is the regime of partially coherent transport of interest here.\cite{Ishizaki2}

There exist two primary design challenges in creating quantum dots assemblies that exhibit partially coherent transport. The first is to reduce the reorganization energy, $\lambda$, as much as possible, taken on in the present investigation. The design of bridge assemblies with sufficiently high excitonic coupling, $J$, will be the subject of a future work. 

The reorganization energy of quantum dots can be reduced by either disrupting or constraining the exciton-phonon interaction because this mediates the structural response to electronic excitation. Each idea is explored with two distinct approaches intended to demonstrate how such mitigation can be designed into quantum dot solids with the aim of enabling partially coherent transport. 

It is not sufficient to simply reduce the reorganization energy, though, as the shape of the spectral density, ${\mathscr J}$, has a strong influence on the rate, $R$, at which the excitonic superposition loses phase coherence. This can be elucidated by considering a SiQD dimer, \cite{Kreisbeck2012, Leonardo} for which this decoherence rate can be expressed as sum of a pure dephasing, $R_d$, and a relaxation, $R_r$:
\begin{eqnarray}
R&=&R_r+R_d \nonumber \\
R_r&\approx& (2 J)^2 S(b)/(2b^2)\label{Rr} \\
R_d&=&\delta \varepsilon^2 S(0)/(2b^2)\label{Rd},
\end{eqnarray}
Here $b^2 = \delta\varepsilon^2+(2J)^2$, $\delta\varepsilon$ is the energy difference between two SiQDs,  $S(\omega) = {\mathscr J}(\omega){\rm coth}(\hbar\omega/(2k_BT))$ is the thermally weighted spectral density, and $k_B$ is Boltzmann's constant.
The dephasing expression determines the rate of decay of off-diagonal elements of the density operator after tracing over all of the phonon states. In contrast, the relaxation expression corresponds to the population transfer between eigenstates of the relevant system that are accompanied by energy dissipation into the phonon reservoir. 

If the two QDs are identical in this dimer system the pure dephasing rate Eq. \ref{Rd} would be zero since $\delta \varepsilon =0$ so that the decoherence rate is determined solely by the relaxation rate of Eq. \ref{Rr}. Within this setting, the rate of decoherence is proportional to the spectral density at a vibrational energy equal to the excitonic coupling, $J$. It is therefore important to suppress, as much as possible, the spectral density for vibrations in this range.

\begin{figure}[hptb]
\begin{center}
\includegraphics[width=0.48\textwidth]{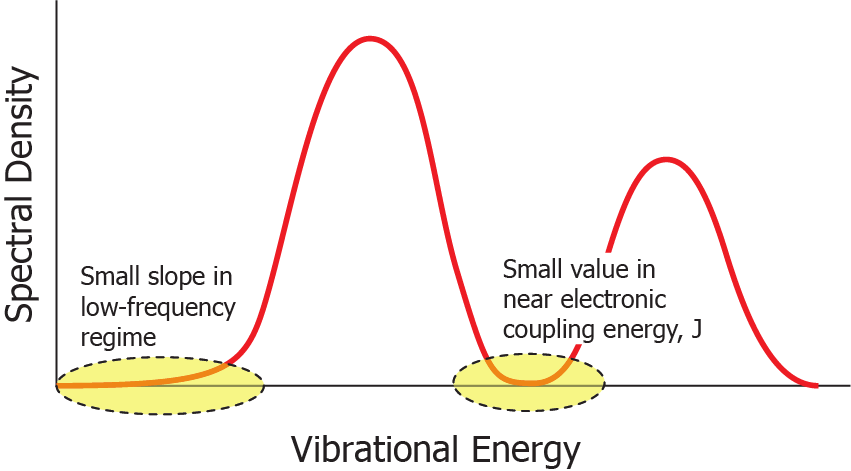}
\end{center}
\caption{Schematic of optimized shape of spectral density, $\mathscr{J}$, in which there exists a small slope for low-energy vibrations and small reorganization energies for vibrational modes in the vicinity of the dot-to-dot electronic coupling, $J$. Along with a low reorganization energy, $\lambda$, high electronic coupling, $J$, spectral densities with this character will lead to longer lifetimes for excitonic superpositions.}
\label{Spectral_Density_Schematic}
\end{figure}

For nonzero $\delta \varepsilon$, on the other hand, the decoherence rate is the sum of rate Eq. \ref{Rd} and Eq. \ref{Rr}. In that case, the pure dephasing rate is proportional to the slope of the spectral density.\cite{Kreisbeck2012}  This implies that a spectral density with a dearth of exciton-phonon coupling at the low-energy end of the vibrational spectrum will result in a flat character there and a small slope at the origin. Both of these key characteristics of the spectral density are illustrated in Figure \ref{Spectral_Density_Schematic}. 
%
%
%

\section{Computational Approach}
All ground  state structures were optimized using density functional theory (DFT) as implemented in DMOL.\cite{Dmol} An all-electron approach was used with exchange and correlation effects accounted for by the generalized gradient approximation (GGA) parameterized by Perdew, Burke and Ernzerhof (PBE)\cite{PBE} for SiQDs and hybrid B3LYP\cite{B3LYP} for organic molecules. A real-space, double numeric plus polarization (DNP) basis was used along with an octopole expansion specify the maximum angular momentum function. Optimization was deemed converged when the energy difference between successive configurations was less than $2\times 10^{-5} \mathrm{Ha}$ and the maximum force on any atom was less than $4\times 10^{-3} \mathrm{Ha}/\mathrm{\AA}$. 

While TDDFT and CI are able to calculate the lowest excitation state and perform geometry relaxation to its equilibrium geometry, these are impractical for large systems since they are too expensive. Instead, a $\Delta$SCF method was employed because it is no more computationally expensive than a small set of ground state DFT calculations. The method is based, in fact, on DFT in which an iterative procedure is used to obtain a set of Kohn-Sham orbitals: 
\begin{equation}
[-\frac{\bigtriangledown^2}{2}+\nu_{ext}(r)+\int\mathrm{d}r^\prime \frac{\rho(r^\prime)}{\lvert r-r^\prime \rvert}+\frac{\delta E_{xc}}{\delta \rho(r)} ]
\label{KS}
\end{equation}
The electron density is the trace of the density operator, constructed using the first N occupied kets:
\begin{equation}
\hat\rho=\sum_{i=1}^{N}\ket{\psi_i}\bra{\psi_i}, \quad \rho = {\rm Tr}(\hat\rho)
\label{DFTden},
\end{equation}

In standard $\Delta$SCF,\cite{Jones,Hellman} a second iterative analysis is carried out but with the electron density operator, $\hat\rho$, now constructed using the lowest $N\mbox{-}1$ orbitals and the $(N\mbox{+}1)^\mathrm{th}$ orbital.  With this change ($\Delta$) imposed on the structure of the density, self-consistent field (SCF) iteration is then carried out on Eq. \ref{KS} as usual, hence the name $\Delta$SCF.  Both accuracy and the rate of convergence can be improved, though, by performing this SCF iteration using a density operator comprised of the lowest $N\mbox{-}1$ orbitals and a linear combination of unoccupied orbitals.\cite{Gavnholt,leDSCF2} This is especially important when there is degeneracy at or near the HOMO or LUMO. Our highly symmetric quantum dots typically exhibit such degeneracies, so we use $1/\sqrt{N_{\rm deg}}\sum_{n_{\rm deg}} \ket{\psi_{n_{\rm deg}}}$ as the $N^\mathrm{th}$ and $(N\mbox{+}1)^\mathrm{th}$ Kohn-Sham orbital, where $N_{\rm deg}$ is the degeneracy of corresponding state. Carrying out iteration process of Eq. \eqref{KS} with this modified density results in the Franck-Condon excitation energy, point A in the left panel of Figure \ref{PES}. To calculate the  reorganization energy, a geometry optimization was subsequently performed to obtain point B in the left panel of Figure \ref{PES}.

This approach does not address the excited state spin, though, which must be antisymmetric (singlet), and some care must be taken to estimate this properly. We first assume that all $N$ orbitals are spin restricted so that the promotion of a Kohn-Sham particle from orbital $N$ to orbital $N\mbox{+}1$ implies that the spin state is determined by particles in orbitals $N\mbox{-}1$ and $N\mbox{+}1$. These two orbitals can then be used to construct the requisite singlet wave function:
\begin{equation}
\ket{\psi_{0}^1}=\frac{1}{\sqrt{2}}(\ket{\uparrow\downarrow} - \ket{\downarrow\uparrow})\label{Singlet}.
\end{equation}
As is typical in electronic structure codes, though, it is not possible to assign such a superposition of spin pairs directly. However, the excited singlet state can also be written as
\begin{equation}
\ket{\psi_0^S}=\sqrt{2}\ket{\uparrow\downarrow} - \ket{\psi_0^T},
\end{equation}
where the triplet spin states are
\begin{eqnarray}
\ket{\psi_1^T}&=&\ket{\uparrow \uparrow}\nonumber\\
\ket{\psi_0^T}&=&\frac{1}{\sqrt{2}}(\ket{\uparrow \downarrow} + \ket{\downarrow \uparrow})\nonumber\\ 
\ket{\psi_{-1}^T}&=&\ket{\downarrow \downarrow}.
\end{eqnarray}
This implies that the desired excited state singlet energy surface, $E_S$, (green curve in both panels of Figure \ref{PES}) can be constructed from the degenerate triplet surface, $E_T$, (red curve in right panel) and that of state $ \ket{\downarrow \uparrow}$ (blue curve in right panel):\cite{SpinPuri}
\begin{equation}
\bra{S}\hat H\ket{S} = E_S = 2E_{\uparrow \downarrow} - E_T .
\label{energy_relation}
\end{equation}
The energy difference between points A and B of on this constructed singlet surface is the reorganization energy, $\lambda$. Since the entirety of this curve is not available in practice, though, we approximate the reorganization energy as the difference between $E_{A'}$ and $E_{B'}$ on the $\ket{\uparrow\downarrow}$ surface (blue) but also calculate  $E_{A"}$ and $E_{B"}$ on the triplet surface (red) for comparison. Note that an analogous approach, building the desired energy surface out of those associated with $\ket{\uparrow \downarrow}$ and $\ket{\downarrow \uparrow}$, will not yield an energy relation analogous to Eq. \ref{energy_relation} which results from one of the constituents being an eigenstate of the Hamiltonian.

The spectral density, $\mathscr{J}$, of Eq. \ref{specdens0} is easily constructed once the reorganization energy, $\lambda$, has been spectrally decomposed into its components, $\lambda_{\xi}$. To implement this, a vibrational (Hessian) analysis is  performed on the ground state structures and the displacement associated with reorganization is expressed as a linear combination of the phonon modes. Anharmonic corrections are also required  for the smallest dots ($\mathrm{Si}_{29}\mathrm{H}_{36}$) but were found to be negligibly small for all larger sizes.

\subsection{Estimated Accuracy of Approach}
The predictions of our computational methodology for calculating reorganization energy can be compared directly with spectral measurements by measuring the Stokes shift--i.e. the energy difference between absorption and emission peaks. The reorganization energy is then half of the Stokes shift. In Table \ref{organic}, predicted reorganization energies for common organic molecules are compared with such experimental measurements with nonpolar  solvents, chosen because the solvent makes a negligible contribution to the reorganization energy. The average discrepancy between $\Delta$SCF and experimental measurement is 84 meV. 
%
%
\begin{table}[hptb]
\caption{Comparison of predicted and measured reorganization energies along with the magnitude of the discrepancy between them. Perylene is in toluene and all other molecules are in cyclohexane. The energy unit is eV. \label{organic}}
\begin{center}
\begin{tabular}{l|ccc}
\hline\hline
& $\Delta$SCF & Exp.&$|$Diff.$|$ \\ \hline
Benzene & 0.26 & 0.18\cite{Berlman}&0.08\\
Naphthalene& 0.32 & 0.30\cite{Berlman}&0.02\\
Anthracene & 0.18 & 0.21\cite{Berlman}&0.03\\
Biphenyl& 0.61& 0.36\cite{Berlman}&0.25\\
Perylene & 0.12 & 0.16\cite{Stopel}&0.04\\
\end{tabular}
\end{center}
\end{table}

While no experimental data is yet available for the reorganization energy of small SiQDs, several computational approaches have been used to predict values as a function of dot size: Quantum Monte Carlo (QMC)\cite{QMC}, Time-Dependent Density Functional Tight-Binding (TDDFTB)\cite{TDDFTB} and Time-Dependent Density Functional Theory (TDDFT)\cite{TDDFT}. It is also common to use a rougher sort of $\Delta$SCF approach wherein ground state DFT orbitals are calculated and then the excited state geometry/energy is obtained via a second DFT calculation in which the ground state HOMO is empty and the LUMO is filled. Table \ref{SiQDdata} shows the comparison of our calculated reorganization energy with these literature results where the rougher $\Delta$SCF scheme is referred to as DFT. With QMC taken to be the most accurate, the average $\Delta$SCF difference from QMC is 57 meV. 

%
\begin{table}[hptb]
\caption{Reorganization energy calculated by $\Delta$SCF, DFT, QMC, TDDFTB and TDDFT. The energy unit is eV. \label{SiQDdata}}
\begin{center}
\begin{tabular}{c|ccccc}
\hline\hline
& $\Delta$SCF& QMC\cite{QMC} & TDDFT\cite{TDDFT}& TDDFTB\cite{TDDFTB}& DFT\cite{QMC}  \\ \hline
$\mathrm{Si}_{29}\mathrm{H}_{36}$ & 0.49 & 0.50& 0.82&0.92 &0.35\\
$\mathrm{Si}_{35}\mathrm{H}_{36}$ & 0.27& 0.40& 0.72&0.74& 0.29\\
$\mathrm{Si}_{66}\mathrm{H}_{64}$& 0.16 &  & &0.52& 0.25\\
$\mathrm{Si}_{29}\mathrm{H}_{24}$& 0.23 & 0.20 & &&0.17  \\
\end{tabular}
\end{center}
\end{table}
%
\section{Designing for Reduced Reorganization Energy}
The reorganization energy of isolated quantum dots can be reduced in one of two ways: by minimizing the affect of exciton creation on structural re-arrangement; and/or by confining the dot so as to inhibit the restructuring that follows excitation. Four design strategies were explored that focus on one or the other of these primary objectives. Mitigation of the affect of the exciton on bonding was explored by: (A1) increasing dot size; and (A2) functionalizing the dot surface so as to modify the frontier orbitals. The influence of structural confinement on reorganization energy was quantified by: (B1) surface reconstruction of the hydrogen bonding; and (B2) encapsulation of the dot within an oxide shell. 

\begin{figure}[hptb]
\begin{center}
\includegraphics[width=0.48\textwidth]{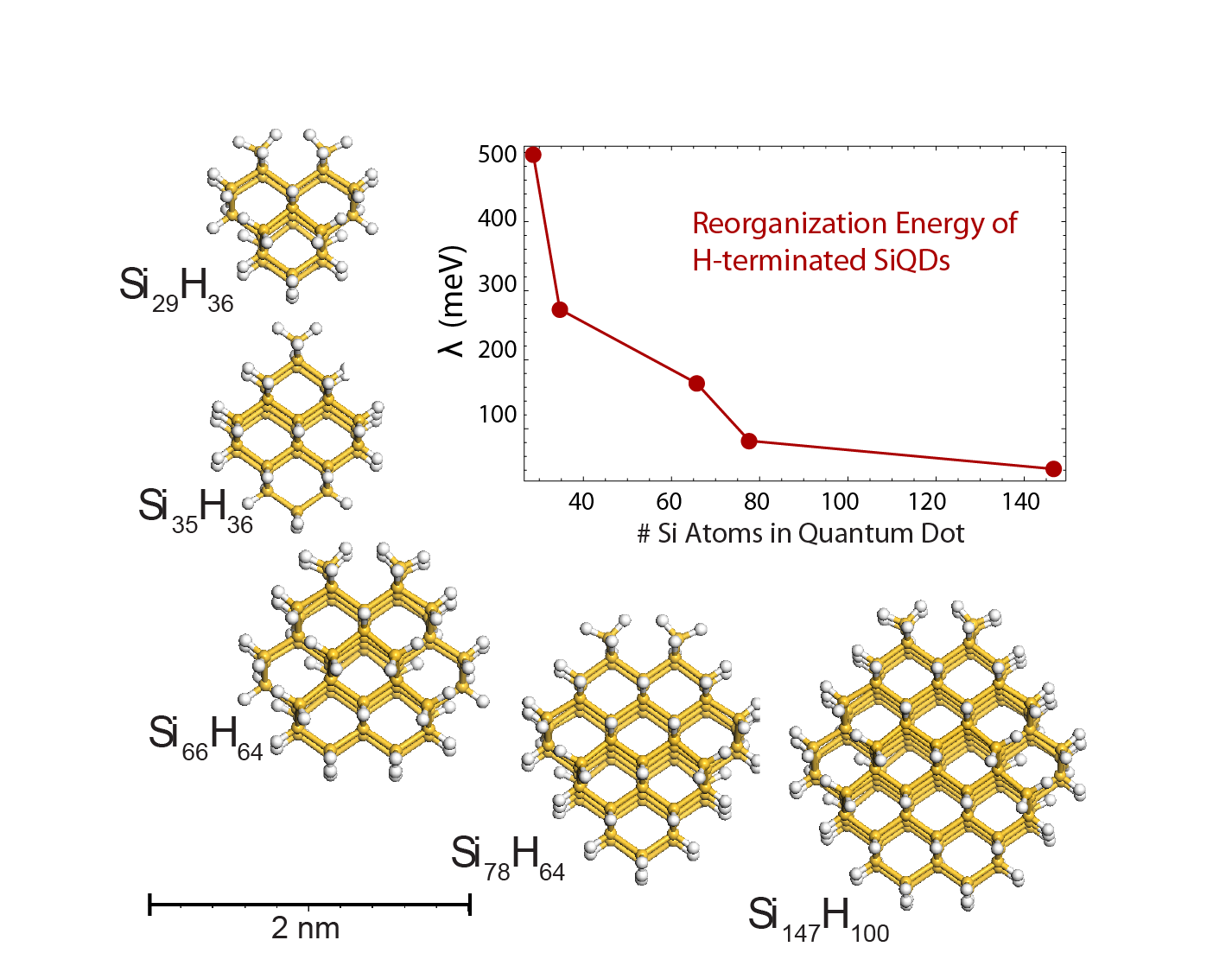}
\end{center}
\caption{The reorganization energy of H-terminated SiQDs changing with number of Si atoms. } \label{ReorgVsNSi}
\end{figure}

\begin{figure}[hptb]
\begin{center}
\includegraphics[width=0.48\textwidth]{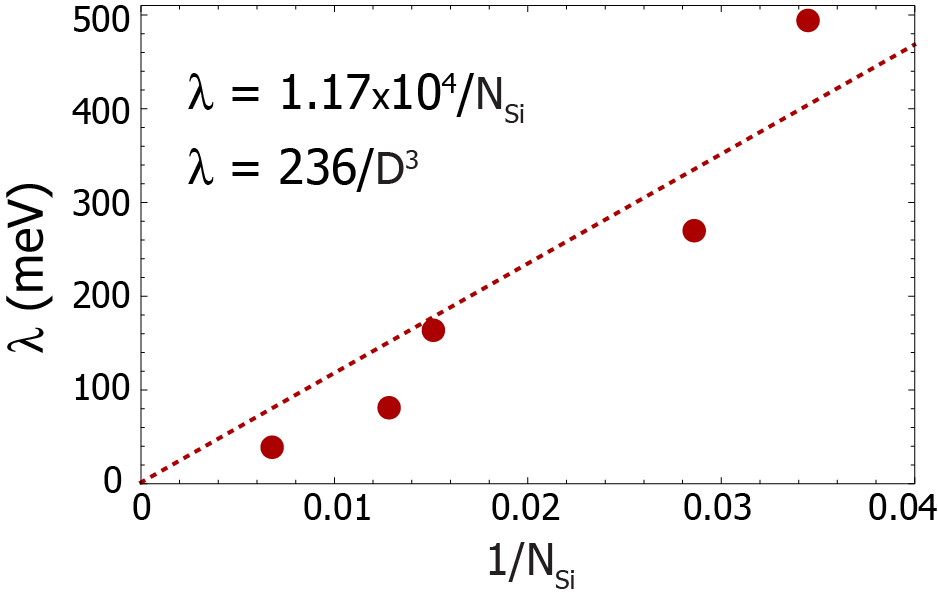}
\end{center}
\caption{The linear dependence of reorganization energy on the inverse of the number of Si atoms per quantum dot, $N_{Si}$. An equivalent relation is also given in terms of the dot diameter, $D$, measured in nm using the bulk diamond Si volume per atom of 0.02 $nm^3$.} \label{reorgVSinvN}
\end{figure}

The most straightforward way of reducing the exciton-phonon coupling is by spreading the exciton out over more atoms. In polyacenes, for instance, the excitonic reorganization energy decreases monotonically with increasing number of phenyl rings.\cite{Klimkans1994} This is due to the extended $\pi$ conjugation of such molecules, but distributed orbitals typify defect-free, crystalline quantum dots as well. Our computational results quantify an analogous trend with increasing dot size, summarized in Figure \ref{ReorgVsNSi} and Table \ref{Reorg_Summary}. It was further found that there exists an inverse linear dependence between excitonic reorganization energy and the number of Si atoms as shown in Figure \ref{reorgVSinvN}. This is consistent with a relationship previously noted for charge transfer excitons in CdSe dots\cite{CX}. As noted in the discussion of computational methods, the reorganization energy as the difference between $E_{A'}$ and $E_{B'}$ on the $\ket{\uparrow\downarrow}$ surface (blue) of the right panel of Fig. \ref{PES}. For comparison, the analogous values associated with $E_{A"}$ and $E_{B"}$ from the triplet surface (red) are:  747 meV ($\mathrm{Si}_{29}\mathrm{H}_{36}$),  317 meV ($\mathrm{Si}_{35}\mathrm{H}_{36}$),  296 meV ($\mathrm{Si}_{66}\mathrm{H}_{64}$) and  82 meV ($\mathrm{Si}_{78}\mathrm{H}_{64}$). From these results it can be concluded that the spin state does not make a meaningful change to the reorganization energy for dots 1.4 nm in diameter or larger.

%
%

%

\begin{figure}[hptb]
\begin{center}
\includegraphics[width=0.44\textwidth]{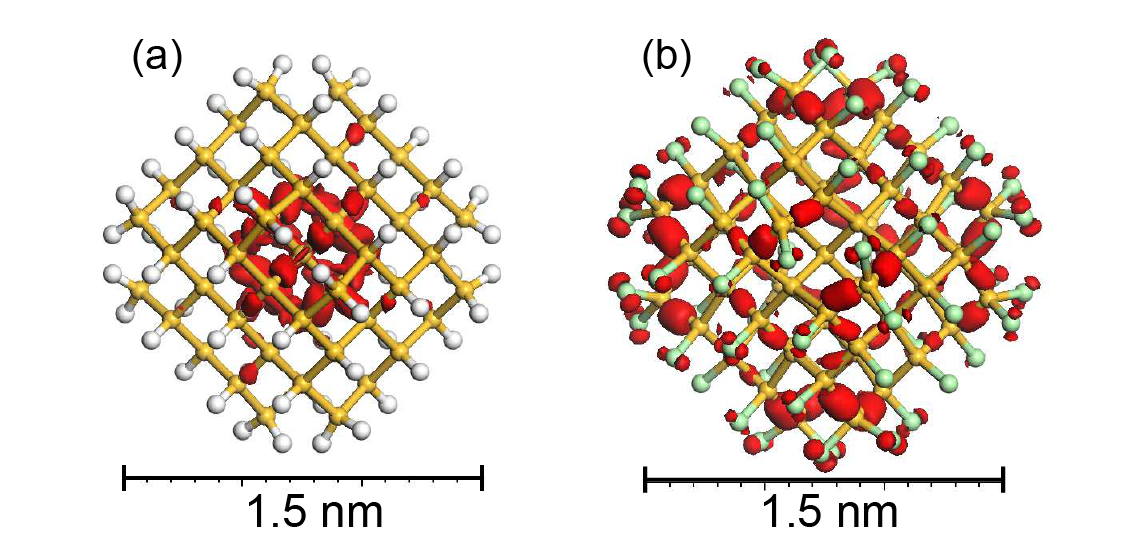}
\end{center}
\caption{ $(a)$ HOMO distribution of H-terminated $\mathrm{Si}_{78}$. $(b)$ HOMO distribution of Cl-terminated $\mathrm{Si}_{78}$. }
\label{Function}
\end{figure}

\begin{figure}[hptb]
\begin{center}
\includegraphics[width=0.48\textwidth]{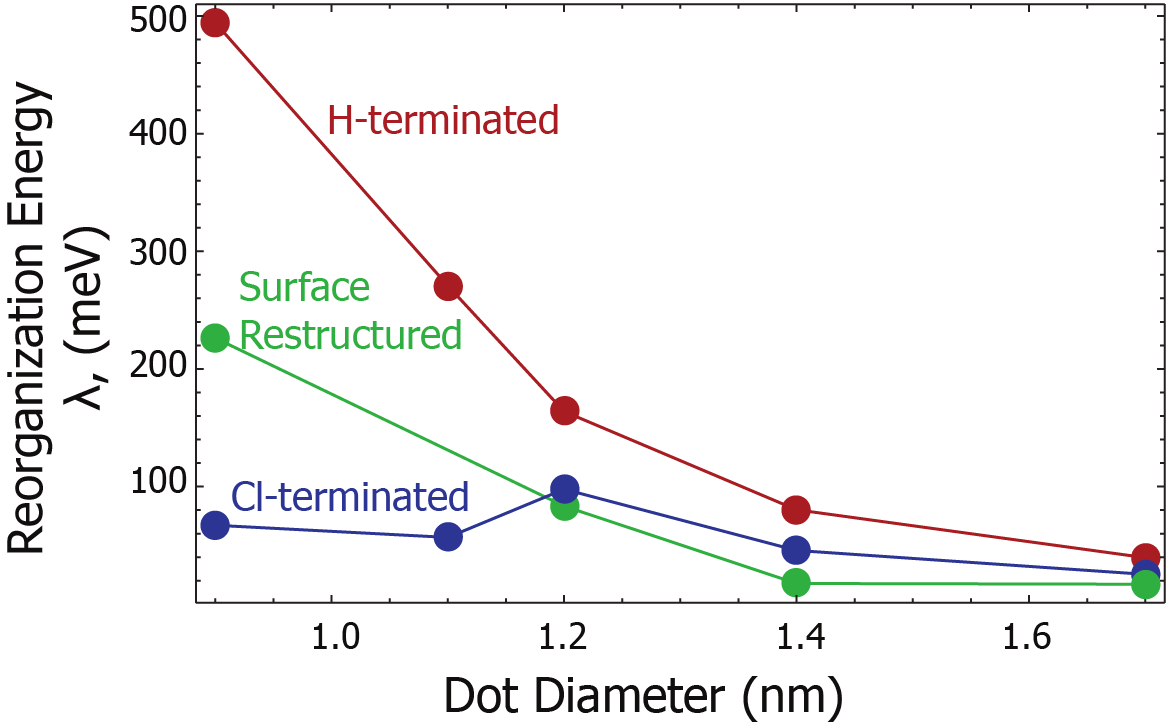}
\end{center}
\caption{Plots of excitonic reorganization energy versus SiQD diameter for: standard H-termination (red); surface reconstructed dots with lower H content (green); and electron-withdrawing Cl-termination (blue). The connecting lines are a guide to the eye.}
\label{Trend}
\end{figure}

Since having a small dot size is important for better transport, fewer defects, and better control of optical properties, consideration was next given as to how to reduce exciton-phonon coupling for dots of fixed size.  This issue has been considered for photosynthetic organic molecules where reorganization energy can be reduced by minimizing the role that frontier orbitals play in bonding. This can be engineered through the addition of electron withdrawing moieties that contribute to the frontier orbitals without significant modification of the bonding orbitals. As a proof of concept, a phenalenyl radical with such properties was found to have record low reorganization energy among organic molecules of similar size.\cite{Chang2010} Such character also underlies the efficiency of lanthanide nanoparticles in carrying out excitonic energy transfer upconversions.\cite{Wang09} There the partially filled 4f level plays the role of frontier orbitals while bonding is dominated by more spatially distributed, filled 5s and 5p states. This strategy can be applied to SiQDs by replacing H-termination with an electron-withdrawing group. Simple Cl termination was deemed sufficient to illustrate the idea, and Figure \ref{Function} shows that, indeed, this functionalization mitigates what had been a strong bonding role played by the HOMO. The LUMO exhibits a similar shift. For the smallest dots, $\mathrm{Si}_{29}\mathrm{H}_{36}$, a drop in reorganization from 494 meV to 67 meV resulted. The results for all dot sizes, provided in Table \ref{Reorg_Summary} and plotted (blue) in Figure \ref{Trend}, indicates that there is not a strong size dependence to this effect. 

%

With two methods explored for reducing the exciton-phonon coupling, we now turn attention to the second design strategy for reducing excitonic reorganization; the application of constraints that prevent or minimize the restructuring despite a driving force to do so. This too has precedence in organic systems. Fullerenes, for instance, have an excitonic reorganization energy that is particularly low because they are extremely rigid.\cite{Andrews2015} This is one of the reasons that they figure so prominently in organic photovoltaic materials such as P3HT/PCBM blends.\cite{PCBM2005}  The excitonic reorganization energy of phthalocyanines, already low, can be further decreased with dimerizing end groups that increase rigidity.\cite{DSouza2009} 

\begin{figure}[hptb]
\begin{center}
\includegraphics[width=0.4\textwidth]{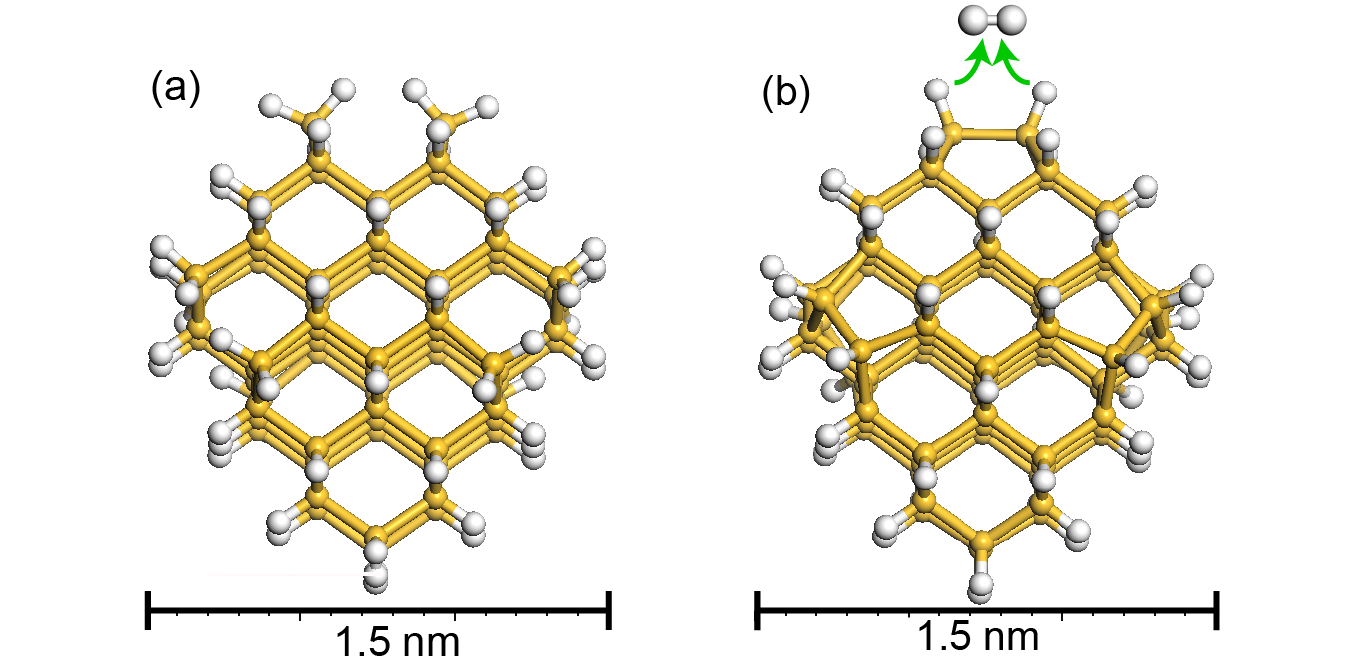}
\end{center}
\caption{(a) $\mathrm{Si}_{78}\mathrm{H}_{64}$ and (b) the same quantum dot after surface restructuring produces 6 $\mathrm{H}_{2}$ molecules leaving $\mathrm{Si}_{78}\mathrm{H}_{52}$.}
\label{Restruct}
\end{figure}

This concept is explored for SiQDs with two types of constraint engineering. The first exploits the fact that dihydride termination of surface silicon atoms can be reduced to monohydride bonding by decreasing the partial pressure of environmental hydrogen.\cite{Gupta1988} This results in additional Si-Si bonding (Figure \ref{Restruct}) that has the structural character of a thin, stiff skin. Such surface reconstruction reduces the reorganization energy as detailed in Table \ref{Reorg_Summary} and plotted (green) in Figure \ref{Trend}. Note that data was not generated for $\mathrm{Si}_{35}\mathrm{H}_{36}$ since the H atoms are not sufficiently close together for surface reconstruction to occur.

%

\begin{figure}[hptb]
\begin{center}
\includegraphics[width=0.48\textwidth]{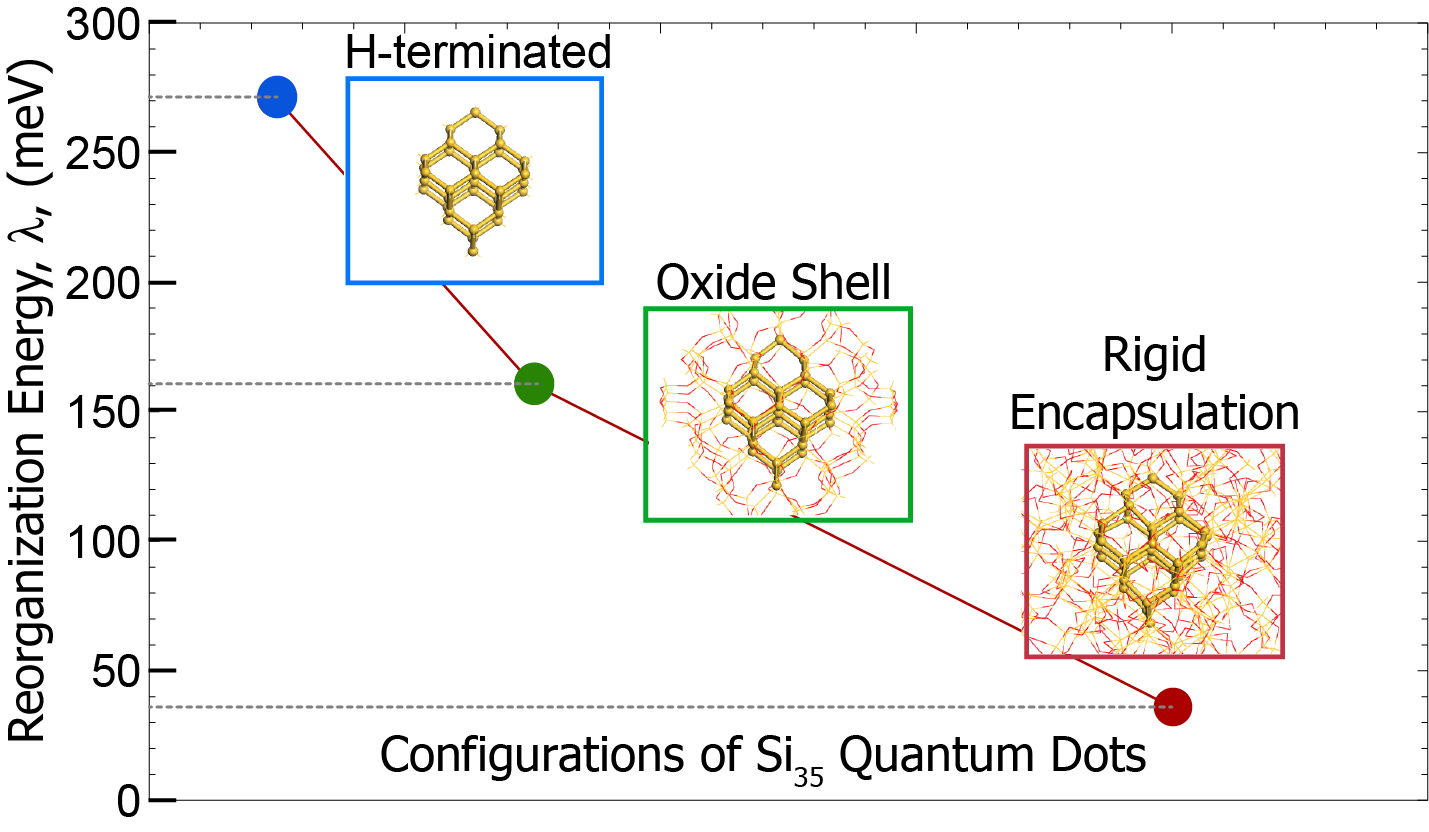}
\end{center}
\caption{ The reorganization trend as $\mathrm{Si}_{35}$ becomes more rigid. The blue dots corresponds to H-terminated $\mathrm{Si}_{35}$; the green dot corresponds to $\mathrm{Si}_{35}$ within one oxidized Si shell; the red dot corresponds to H-terminated $\mathrm{Si}_{35}$ with H fixed to mimic $\mathrm{Si}_{35}$ within an infinite large matrix.}
\label{Encapsulation}
\end{figure}

A second approach for constraining structural rearrangement is to encapsulate SiQDs within a rigid matrix. To test this idea, an $\mathrm{Si}_{35}$ dot was encapsulated within a monolayer of $\mathrm{SiO_2}$. As shown in the center of Figure \ref{Encapsulation}, this reduces the excitonic reorganization energy from 271 meV to 161 meV. To develop a sense for the impact a thicker oxide shell would make, we also calculated the reorganization energy for  $\mathrm{Si}_{35}\mathrm{H}_{36}$ with the H atoms held fixed. This resulted in a remarkably low excitonic reorganization energy of 36 meV, as shown at right in Figure \ref{Encapsulation}.

%
%

A summary of all four analyses is provided in Table \ref{SiQDs} and Figure \ref{Trend}.  The lowest reorganization energy achieved is 17 meV (137 $\mathrm{cm}^{-1}$) for a surface reconstructed SiQD with a diameter of 1.7 nm.

%
\begin{table}[hptb]
\caption{Reorganization energy of H-terminated, Cl-terminated and surface-restructured SiQDs. The energy unit is meV. \label{SiQDs}}
\begin{center}
\begin{tabular}{l|ccccc}
\hline\hline
SiQD diameter(nm) & 0.9 & 1.1 & 1.2 & 1.4 & 1.7 \\ \hline
\# Si Atoms & 29 & 35 & 66 & 78 & 147 \\ \hline
H-terminated & 494 & 271 & 164 & 80&40\\
Cl-terminated& 67 & 57 & 98 & 46& 25\\
Restructured & 226 & $\mathrm{-}$ & 83 & 18 & 17\\
\end{tabular}
\label{Reorg_Summary}
\end{center}
\end{table}
%
%

\begin{figure}[hptb]
\begin{center}
\includegraphics[width=0.48\textwidth]{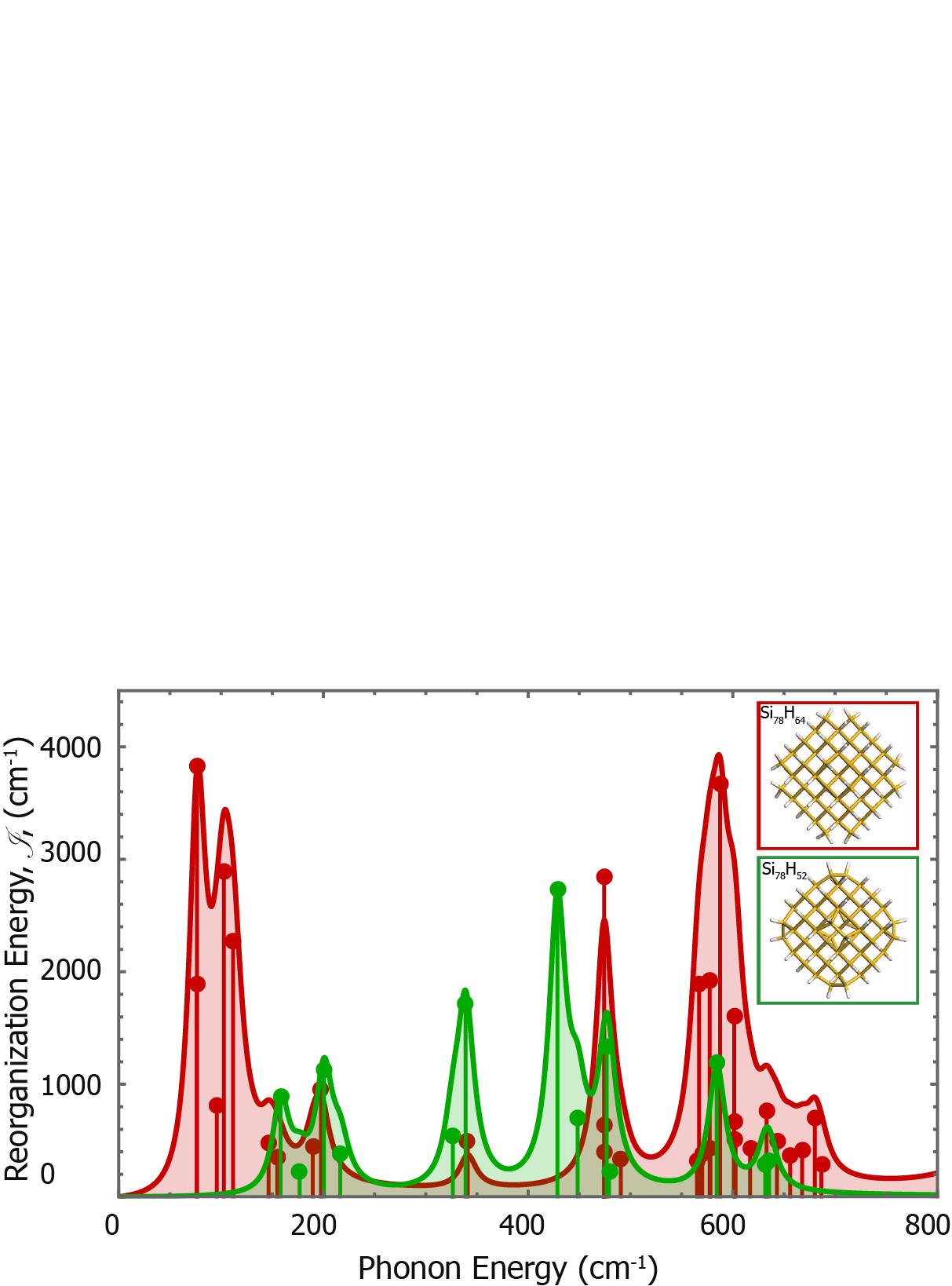}
\end{center}
\caption{Spectral density of ${\rm Si}_{78}{\rm H}_{64}$ (red) compared with that of its surface-restructured counterpart, ${\rm Si}_{78}{\rm H}_{52}$ (green). Vertical lines indicate non-negligible contributions to the reorganization energy and have been normalized to the peak value of the fitted spectral density.} 
\label{Spectra}
\end{figure}

We now turn to a consideration of the shape of the spectral density to determine the extent to which it can be optimized along the lines of Figure \ref{Spectral_Density_Schematic}. As a specific example, the spectral density of ${\rm Si}_{78}{\rm H}_{64}$ dots is compared with that of its restructured counterpart, ${\rm Si}_{78}{\rm H}_{52}$.  Figure \ref{Spectra} shows both spectra along with their Drude-Lorentz fits using Eq. \ref{specdens} with 34 peaks for ${\rm Si}_{78}{\rm H}_{64}$ and 13 peaks for ${\rm Si}_{78}{\rm H}_{52}$.  Reconstruction modifies the lower energy end of the spectrum in a way that should significantly reduce the decoherence rate, and restructuring  also suppresses the spectrum in the physically desirable 500--700 $\mathrm{cm}^{-1}$ range of electronic couplings. This too would lead to a system more capable of supporting partially coherent transport.  

%
%
%

\section{Conclusions}
Sufficiently low dielectric screening in quantum dot assemblies makes charge dissociation improbable following photo excitation. Energy is then transported via excitons in what is typically a random walk. Such diffusive propagation is the result of a loss of coherence in the superposition of excitonic states that are present in the initial photo-absorption, but maintaining an element of such coherence is thought to result in a robust, more efficient form of energy transport. This requires a low reorganization energy for isolated dots and a large excitonic coupling between dots. The current work has focused on ways in which reorganization energy can be reduced by either reducing or constraining the associated exciton-phonon coupling. Two design paradigms were explored for each of these. 

Reduction in the coupling itself can be accomplished most simply by spreading out the electronic excitation over more atoms, for instance by increasing dot size.  The reorganization energy, $\lambda$, was found to be inversely proportional to the cube of the dot diameter, $D$, and a simple relationship was identified: $\lambda = 236/D^3$ with $D$ in nm and $\lambda$ in meV. Since having a small dot size is important for better transport, fewer defects and easier functionalization, though, a second way of reducing exciton-phonon coupling was also explored. Dots were functionalized so that the frontier orbitals, of which excitons are primarily comprised, are not the bonding orbitals that generate phonons. This idea has been successfully demonstrated for small organic molecules, and it is also underlies the efficiency of lanthanide nanoparticles in efficiently carrying out excitonic energy transfer upconversions.\cite{Wang09} Our analysis indicates that this is also possible for SiQDs using Cl atoms--not because Cl is considered a candidate functionalization but to demonstrate proof of the concept. For the smallest dots, a drop in reorganization from 494 meV to 67 meV resulted. This is a surface effect, however, and further reduction in reorganization energy is minimal as dot size is increased.

The second design paradigm is to constrain the system so that it cannot carry out the structural re-arrangements that generate reorganization energy. This was first explored by controlling the hydrogenation of dot surfaces which can be carried out by changing the partial pressure of hydrogen in the environment.\cite{Gupta1988} The results indicate the reorganization energy is significantly reduced as a result of what amounts to a stiff skin on the surface of the dot that resists the driving force to reorganize. As with the standard H-terminated dots, this reorganization energy decreases with increasing dot size, and 17 meV was the minimum value computationally measured. Matrix encapsulation is also a promising avenue for reducing the motion that generates the reorganization energy well. A monolayer of ${\rm SiO}_{2}$ around ${\rm Si}_{35}$ reduced $\lambda$ from 271 meV to 161 meV and freezing the H-atoms on the dot surface dropped this value down to only 36 meV.

This computational study of 1-2 nm silicon quantum dots is not focused on reducing excitonic reorganization energy as much as possible. Rather, it identifies ways in which this energy can be reduced in order to offer direction to future work. Even so, the results suggest that it is reasonable to identify a goal of creating 1-2 nm silicon quantum dots with reorganization values in the range of 1-10 meV. This overlays the range typically considered in association with photosynthetic complexes of 0.1 to 10 meV\cite{IshizakiPNAS2009}--naturally occurring assemblies that serve as a bio-inspiration for what may be improved upon with engineered organic/inorganic nanostructures.\cite{Engel1, Adolphs4, Vulto5,Cho6, Brixner7} More important than the range itself, though, successful designs will need to produce excitonic coupling of the same order or larger to create quantum dot solids with a measure of wave-like character.

\section{Acknowledgements}
We are grateful to C. Kreisbeck and A. Aspuru-Guzik for discussions of reorganization energy in organic systems. This work was supported by the Renewable Energy Materials Research Science and Engineering Center (NSF Grant No. DMR-0820518) at the Colorado School of Mines. The calculations were carried out using the high performance computing resources provided by the Golden Energy Computing Organization at the Colorado School of Mines (NSF Grant No. CNS-0722415).


\end{document}